\expandafter\ifx\csname LaTeX\endcsname\relax
      \let\maybe\relax
\else \immediate\write0{}
      \message{You need to run TeX for this, not LaTeX}
      \immediate\write0{}
      \makeatletter\let\maybe\@@end
\fi
\maybe

\magnification=\magstephalf

\hsize=5.25truein
\vsize=8.3truein
\hoffset=0.37truein

\newdimen\frontindent \frontindent=.45truein
\newdimen\theparindent \theparindent=20pt


\let\em=\it

\font\tencsc=cmcsc10
\font\twelvebf=cmbx10 scaled 1200
\font\bmit=cmmib10  \font\twelvebmit=cmmib10 scaled 1200
\font\sixrm=cmr6 \font\sixi=cmmi6 \font\sixit=cmti8 at 6pt

\font\eightrm=cmr8  \let\smallrm=\eightrm
\font\eighti=cmmi8  \let\smalli=\eighti
\skewchar\eighti='177
\font\eightsy=cmsy8
\skewchar\eightsy='60
\font\eightit=cmti8
\font\eightsl=cmsl8
\font\eightbf=cmbx8
\font\eighttt=cmtt8
\def\eightpoint{\textfont0=\eightrm \scriptfont0=\fiverm 
                \def\rm{\fam0\eightrm}\relax
                \textfont1=\eighti \scriptfont1=\fivei 
                \def\mit{\fam1}\def\oldstyle{\fam1\eighti}\relax
                \textfont2=\eightsy \scriptfont2=\fivesy 
                \def\cal{\fam2}\relax
                \textfont3=\tenex \scriptfont3=\tenex 
                \def\it{\fam\itfam\eightit}\let\em=\it
                \textfont\itfam=\eightit
                \def\sl{\fam\slfam\eightsl}\relax
                \textfont\slfam=\eightsl
                \def\bf{\fam\bffam\eightbf}\relax
                \textfont\bffam=\eightbf \scriptfont\bffam=\fivebf
                \def\tt{\fam\ttfam\eighttt}\relax
                \textfont\ttfam=\eighttt
                \setbox\strutbox=\hbox{\vrule
                     height7pt depth2pt width0pt}\baselineskip=9pt
                \let\smallrm=\sixrm \let\smalli=\sixi
                \rm}


\catcode`@=11
 
\def\vfootnote#1{\insert\footins\bgroup\eightpoint
     \interlinepenalty=\interfootnotelinepenalty
     \splittopskip=\ht\strutbox \splitmaxdepth=\dp\strutbox
     \floatingpenalty=20000
     \leftskip=0pt \rightskip=0pt \parskip=1pt \spaceskip=0pt \xspaceskip=0pt
     \everydisplay={}
     \smallskip\textindent{#1}\footstrut\futurelet\next\fo@t}
 
\newcount\notenum

\def\note{\global\advance\notenum by 1
    \edef\n@tenum{$^{\the\notenum}$}\let\@sf=\empty
    \ifhmode\edef\@sf{\spacefactor=\the\spacefactor}\/\fi
    \n@tenum\@sf\vfootnote{\n@tenum}}


\tabskip1em

\newtoks\pream \pream={#\strut}
\newtoks\lpream \lpream={&#\hfil}
\newtoks\rpream \rpream={&\hfil#}
\newtoks\cpream \cpream={&\hfil#\hfil}
\newtoks\mpream \mpream={&&\hfil#\hfil}

\newcount\ncol \def\ncolp{\advance\ncol by 1}
\def\atalias#1{
    \ifx#1l\edef\xpream{\pream={\the\pream\the\lpream}}\xpream\ncolp\fi
    \ifx#1r\edef\xpream{\pream={\the\pream\the\rpream}}\xpream\ncolp\fi
    \ifx#1c\edef\xpream{\pream={\the\pream\the\cpream}}\xpream\ncolp\fi}
\catcode`@=\active

\def\taborl#1{\omit\unskip#1\hfil}
\def\taborc#1{\omit\hfil#1\hfil}
\def\taborr#1{\omit\hfil#1}
\def\multicol#1{\multispan#1\let\omit\relax}

\def\table#1\par{\midinsert\offinterlineskip\everydisplay{}
    \let@\atalias \let\l\taborl \let\r\taborr \let\c\taborc
    \def\space{\noalign{\vskip2pt}}
    \def\tablerule{\omit&\multispan{\the\ncol}\hrulefill\cr}
    \def\onerule{\space\space\tablerule\space\space}
    \def\tworules{\space\space\tablerule\space\tablerule\space\space}
    \def\annot##1\\{&\multispan{\the\ncol}##1\hfil\cr}
    \def\\{\let\\=\cr
           \edef\xpream{\pream={\the\pream\the\mpream}}\xpream
           \edef\starthalign{$$\vbox\bgroup\halign\bgroup\the\pream\cr}
           \starthalign
           \annot\hfil\tencsc Table #1\\ \noalign{\medskip}}
    \let\par\endtable}

\edef\endtable{\noalign{\vskip-\bigskipamount}\egroup\egroup$$\endinsert}

\let\plainmidinsert=\midinsert
\def\eightpttable{\def\midinsert{\let\midinsert=\plainmidinsert
    \plainmidinsert\eightpoint\tabskip 1em}\table}



\newif\iftitlepage

\def\raggedright{\rightskip 0pt plus .2\hsize\relax}

\let\caret=^ \catcode`\^=13 \def^#1{\ifmmode\caret{#1}\else$\caret{#1}$\fi}

\def\title#1\par{\vfill\supereject\begingroup
                 \global\titlepagetrue
                 \leftskip=\frontindent\parindent=0pt\parskip=0pt
                 \frenchspacing \eqnum=0
                 \gdef\runningtitle{#1}
                 \null\vskip-22.5pt\copy\volbox\vskip18pt
                 {\titlestyle#1\bigskip}}
\def\titlestyle{\raggedright\bf\twelvebf\textfont1=\twelvebmit
                \let\smallrm=\tenbf \let\smalli=\bmit
                \baselineskip=1.2\baselineskip}
\def\shorttitle#1\par{\gdef\runningtitle{#1}}
\def\author#1\par{{\raggedright#1\medskip}}

\def\shortauthor#1\par{\gdef\runningauthors{#1}}

\def\affil#1\par{{\raggedright\it#1\smallskip}}
\def@#1{\ifhmode\qquad\fi\leavevmode\llap{^{#1}}\ignorespaces}
\def\abstract{\smallskip\medskip{\bf Abstract: }}

\def\maybebreak#1{\vskip0pt plus #1\hsize \penalty-500
                  \vskip0pt plus -#1\hsize}

\def\maintextmode{\leftskip=0pt\parindent=\theparindent
                  \parskip=\smallskipamount\nonfrenchspacing}

\def\maintext#1\par{\bigskip\medskip\maintextmode\noindent}

\newcount\secnum
\def\section#1\par{\ifnum\secnum=0\medskip\maintextmode\fi
    \advance\secnum by 1 \bigskip\maybebreak{.1}
    \subsecnum=0
    \hang\noindent\hbox to \parindent{\bf\the\secnum.\hfil}{\bf#1}
    \smallskip\noindent}

\newcount\subsecnum
\def\subsection#1\par{\ifnum\subsecnum>0\medskip\maybebreak{.1}\fi
    \advance\subsecnum by 1
    \hang\noindent\hbox to \parindent
       {\it\the\secnum.\the\subsecnum\hfil}{\it#1}
    \par\noindent}

\def\references\par{\bigskip\maybebreak{.1}\parindent=0pt
    \everypar{\hangindent\theparindent\hangafter1}
    \leftline{\bf References}\smallskip}

\def\appendix#1\par{\bigskip\maybebreak{.1}\maintextmode
    \advance\secnum by 1 \bigskip\maybebreak{.1}
    \leftline{\bf Appendix: #1}\smallskip\noindent}

\def\acknowl{\medskip\noindent}

\def\bye{\endgroup\vfill\supereject\end}


\newbox\volbox
\setbox\volbox=\vbox{\hsize=.5\hsize \raggedright
       \sixit\baselineskip=7.2pt \noindent
       Nonlinear Dynamics in Astronomy and Physics,
       A Workshop Dedicated to the Memory of Professor
       Henry E. Kandrup.
       To appear in Annals of the New York Academy of Sciences,
       Eds.\ J.R.~Buchler, S. T. Gottesman, M. E. Mahon}


\input epsf

\def\figureps[#1,#2]#3.{\midinsert\parindent=0pt\eightpoint
    \vbox{\epsfxsize=#2\centerline{\epsfbox{#1}}}
    \def\par{\endgraf\endinsert}{\bf Figure#3.}}

\def\figuretwops[#1,#2,#3]#4.{\midinsert\parindent=0pt\eightpoint
    \vbox{\centerline{\epsfxsize=#3\epsfbox{#1}\hfil
                      \epsfxsize=#3\epsfbox{#2}}}
     \def\par{\endgraf\endinsert}{\bf Figure#4.}}

\def\figurespace[#1]#2.{\midinsert\parindent=0pt\eightpoint
    \vbox to #1 {\vfil\centerline{\tenit Stick Figure#2 here!}\vfil}
    \def\par{\endgraf\endinsert}{\bf Figure#2.}}


\output={\plainoutput\global\titlepagefalse}


\newcount\eqnum
\everydisplay{\puteqnum}  
\def\puteqnum#1$${#1\global\advance\eqnum by 1\eqno(\the\eqnum)$$}
\def\namethiseqn#1{\xdef#1{\the\eqnum}}

 
\newcount\mpageno
\mpageno=\pageno  \advance\mpageno by 1000
 
\def\advancepageno{\global\advance\pageno by 1
                   \global\advance\mpageno by 1 }

\openout15=inx
\def\index#1{\write15{{#1}{\the\mpageno}}\ignorespaces}


\def\LaTeX{{\rm L\kern-.36em\raise.3ex\hbox{\tencsc a}\kern-.15em
    T\kern-.1667em\lower.7ex\hbox{E}\kern-.125emX}}

\def\[#1]{\raise.2ex\hbox{[}#1\raise.2ex\hbox{]}}

\def\witchbox#1#2#3{\hbox{$\mathchar"#1#2#3$}}
\def\leqsim{\mathrel{\rlap{\lower3pt\witchbox218}\raise2pt\witchbox13C}}
\def\geqsim{\mathrel{\rlap{\lower3pt\witchbox218}\raise2pt\witchbox13E}}

\def\<#1>{\langle#1\rangle}


{\obeyspaces\gdef {\ }}

\catcode`@=12 \let\@=@ \catcode`@=13
\def\+{\catcode`\\=12\catcode`\$=12\catcode`\&=12\catcode`\#=12%
       \catcode`\^=12\catcode`\_=12\catcode`\~=12\catcode`\%=12%
       \catcode`\@=0\tt}
\def\({\endgraf\bgroup\let\par=\endgraf\parskip=0pt\vskip3pt
       \eightpoint \def\/{{\eightpoint$\langle$Blank line$\rangle$}}
       \catcode`\{=12\catcode`\}=12\+\obeylines\obeyspaces}
\def\){\vskip1pt\egroup\vskip-\parskip\noindent\ignorespaces}


\def\ltsima{$\; \buildrel < \over \sim \;$}
\def\simlt{\lower.5ex\hbox{\ltsima}}
\def\gtsima{$\; \buildrel > \over \sim \;$}
\def\simgt{\lower.5ex\hbox{\gtsima}}

\title EVOLUTION OF BINARY SUPERMASSIVE BLACK HOLES VIA CHAIN REGULARIZATION

\shorttitle Binary Black Holes

\author Andras Szell and David Merritt 

\shortauthor Szell, Merritt \& Mikkola

\affil Rochester Institute of Technology

\medskip

\author Seppo Mikkola

\affil Tuorla Observatory

\bigskip

\abstract
A chain regularization method is combined with special purpose
computer hardware to study the evolution of massive black hole
binaries at the centers of galaxies.
Preliminary results with up to $N=0.26\times 10^6$ particles
are presented.
The decay rate of the binary is shown to decrease
with increasing $N$, as expected on the basis of
theoretical arguments.
The eccentricity of the binary remains small.

\bigskip\maintextmode

Coalescence of binary supermassive black holes 
is potentially the strongest source of gravitational
waves in the universe [1].
The coalescence rate is limited by the efficiency
with which massive binaries can interact with
stars and gas in a galaxy and reach the relativistic
regime at separations of $\sim 10^{-3}$ pc.
Exchange of energy between a binary black hole and 
stars should also leave observable traces in the stellar 
distribution, perhaps allowing us
to infer something about the merger history of
galaxies from their nuclear structure [2].
Henry Kandrup worked on this problem shortly before
his death.
In ``Supermassive Black Hole Binaries as Galactic Blenders''
[3], Kandrup {\it et al.} investigated the effects of
a massive binary on the stellar orbits near the
center of spherical and nearly spherical galaxies.
They showed that the periodically-varying potential
due to the binary, coupled with the fixed potential
from the galaxy, was effective at inducing chaos
in the stellar orbits, leading to diffusion in both
energy and configuration space and to ejection
of stars from the nucleus. 
This study was a complement to earlier studies
based on scattering experiments [4,5] in which
the potential of the galaxy was ignored.

Another approach to the binary black hole problem
is via direct $N$-body techniques [6-8].
This approach is computationally challenging
because of the need to handle close interactions
between the star- and black hole particles with high precision.
In addition, large particle numbers are required
to avoid the effects of spurious relaxation [9,10].
Here, we present preliminary results of $N$-body
integrations of the binary black hole problem,
in which close interactions between the black holes
and stars are handled via the Mikkola-Aarseth
chain-regularization algorithm [11,12].
Recently Aarseth [13] described an application of a 
time transformed leapfrog scheme [14] to this problem. 
We prefer the chain algorithm since it has proved 
itself in numerous applications, including one
very similar to the current problem [15].
We incorporate the chain algorithm into a
general-purpose $N$-body code by including the
effects of nearby stars as perturbers to the chain.
We present some preliminary results of binary black hole
evolution computed via this algorithm on a special-purpose
GRAPE-6 computer with particle numbers up to $0.26\times 10^6$.

Our basic $N$-body algorithm is an adaptation of the {\tt NBODY1}
code of Aarseth [16] to the GRAPE-6 special purpose hardware.
The code uses a fourth-order Hermite integration scheme
with individual, adaptive, block time steps [17].
For the majority of the particles, the forces and force
derivatives were calculated via a direct-summation scheme using
the GRAPE-6.

Close encounters between the massive particles
(``black holes''), or between black holes and stars,
create prohibitively small time steps in such a scheme.
To avoid this situation, we regularized the critical
interactions as follows.
Let ${\bf r}_i$, $i=1,...,N$ be the position vectors of the particles.
We first identify the subset of $n$ particles to be included
in the chain; the precise criterion for inclusion is
presented below,
but in the late stages of evolution, the chain always
included the two black holes as its lowest members.
We then search for the particle which is closest to either end
of the chain and add it; this operation is repeated recursively
until all $n$ particles are included.
Define the separation vectors
${\bf R}_i = {\bf r}_{i+1} - {\bf r}_i$
where ${\bf r}_{i+1}$ and ${\bf r}_i$ are the coordinates
of the two particles making up the $i$th link of the chain.
The canonical momenta ${\bf W}_i$ corresponding to the
coordinates ${\bf R}_i$ are given in terms of the old momenta
via the generating function
$$
	S = \sum_{i=1}^{n-1} {\bf W}_i\cdot ({\bf r}_{i+1} - {\bf r}_i).
$$
Next, we apply KS regularization [18] to the chain vectors,
regularizing only the interactions between neighboring
particles in the chain.
Let ${\bf Q}_i$ and ${\bf P}_i$ be the KS transformed ${\bf R}_i$ and
${\bf W}_i$ coordinates.
After applying the time transormation $\delta t= g \delta s$,
$g = 1/L$, where $L$ is the Lagrangian of the system
($L = T - U$, where $T$ is the kinetic and $U$ is the
potential energy of the system).
 We obtain the regularized Hamiltonian
$\Gamma = g (H({\bf Q}_i,{\bf P}_i) - E_0)$,
where $E_0$ is the total energy of the system.
The equations of motion are then
$$
{\bf P}_i' =
- {\partial {\bf \Gamma}\over \partial {\bf Q}_i} \;,
\quad
{\bf Q}_i' = {\partial {\bf \Gamma}\over \partial {\bf P}_i}
$$
where primes denote differentiation with respect to the time
coordinate $s$.
Because of the use of regularized coordinates, these
equations do not suffer from singularities, 
as long as care is taken in the construction of the chain.

Since it is impractical to include all $N$ particles in the
chain, we must consider the effects of external
forces on the chain members.
Let ${\bf F}_j$ be the perturbing acceleration acting on 
the $j$th body of mass $m_j$.
The perturbed system can be written in Hamiltonian form by 
simply adding the perturbing potential:
$$
\delta U = \sum_{j=1}^n m_j {\bf r}_j \cdot {{\bf F}_j}(t).
$$

Only one chain was defined at any given time.
At the start of the $N$-body integrations,
there was no regularization, and all particles were
advanced using the variable-time-step Hermite scheme.
The chain was ``turned on'' at the time when one of
the particles (including possibly one of the black holes)
achieved a time step shorter than $t_{chmin}$ and
reached a distance from one of the black holes smaller
than $r_{chmin}$.
Each star inside $r_{chmin}$ radius was then added to the chain,
and the two black holes were always included.
The values of $t_{chmin}$ and $r_{chmin}$ 
were determined by carrying out test runs; we adopted
$t_{chmin} \approx 10^{-5} - 10^{-6}$ and $r_{chmin} \approx 10^{-4} - 10^{-3}$
in standard $N$-body units.

The chain's center of mass was a pseudoparticle as seen by the 
$N$-body code and was advanced by the Hermite scheme in the same 
way as an ordinary particle. 
However, when integrating the trajectories of stars near to the chain,
it is essential to resolve the inner structure of the chain.
Thus for stars inside a critical $r_{crit1}$ radius around the
chain, the forces from the individual chain members were taken 
into account. 
The value of $r_{crit1}$ was set by the size of the chain to be
$r_{crit1} = \lambda R_{ch}$ with $R_{ch}$ the spatial size 
of the chain and $\lambda = 100$.

In addition, the equations of motion of the chain particles 
must include the forces exerted by a set of external perturber stars. 
Whether or not a given star was listed as a perturber was determined by
a tidal criterion: 
$r < R_{crit2} = (m/m_{chain})^{1/3} \gamma_{min}^{-1/3} R_{ch}$ 
where $m_{chain}$ represents the mass of the chain, 
$m$ is the mass of the star, and $\gamma_{min}$ was chosen to be $10^{-6}$;
 thus $r_{crit2} \approx 10^{2} (m/m_{chain})^{1/3}  R_{ch}$.

The membership of the chain changed under the evolution of the system. 
Stars were captured into the chain if their orbits approached the binary 
closer than $R_{ch}$. Stars were emitted from the
chain if they got further from both of the black holes than 
$1.5 R_{ch}$. The difference between the emission
and absorption distances was chosen to avoid a too-frequent 
variation of the chain membership.
When the last particle left the chain, 
the chain was eliminated and the integration turned back
to the Hermite scheme, until a new chain was created.

In our numerical experiments, the typical number of chain members 
was $5 - 10$. 
The number of perturber stars was typically $500-1000$,
and the number of stars inside $r_{crit1}$ was $2000 - 5000$.
Using a minimum tolerance of $10^{-12}$ for the chain's Bulirsch-Stoer 
integrator allowed us to reach typical relative accuracies
of $10^{-9}$ in the chain integration. 
The relative accuracy in the conservation of the total energy of the system
was determined by the Hermite scheme. 
For all of our numerical simulation the relative error in the total 
energy was less than $10^{-4}$ during the course of the integration.

For this set of experiments, we adopted Dehnen's [19] density law,
$$
\rho (r) = {(3 - \gamma) M \over 4 \pi} {a \over r^{\gamma}(r + a)^{4 - \gamma}} \;,
$$
with $\gamma=0.5$, to describe the initial stellar distribution.
The initial positions and velocities of the $N$ stars were generated 
from the steady state phase space density $f(E)$ that reproduces 
the density law (4).
Henceforth we adopt units such that the gravitational
constant $G$, the total stellar mass $M$, and the Dehnen scale length
$a$ are equal to one.
To  this model we added two black hole particles each of mass
$0.005$.
The black holes were placed symmetrically about the center of the
galaxy, offset at a distance $0.1$ from the center.
The tangential velocities were set to $\pm 0.16$
yielding nearly circular initial
orbits for the two black holes about the center of the galaxy.

We integrated the above model with three different particle numbers:
$N = 16384$, $65536$, and $262144$; the latter is close to
the maximum number of particles that can be handled in the GRAPE-6
memory.
The integrations were carried out for $170$ time units.
Elapsed times were ($2.5, 26, 105$) hours for the three runs.

\figureps[f1.ps,0.7\hsize]{\ 1.}
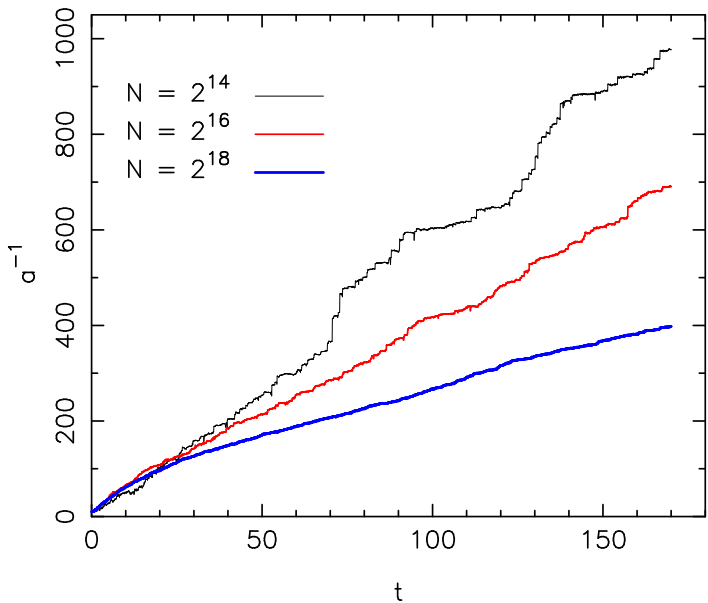
Evolution of the binary semi-major axis.

The orbits of the two black holes initially decay,
and at a time of roughly $30$ they form a bound pair.
After this, the semimajor axis $a$ of the binary
shrinks as the two black holes interact
with stars and eject them from the nucleus via the
gravitational slingshot.
The instantaneous decay rate is given approximately by
$$
{d\over dt}\left({1\over a}\right) = {G\rho\over\sigma}H
$$
where $\rho$ and $\sigma$ are the density and velocity dispersion
of the stars, and $H$ is a dimensionless constant of
order $16$ [4,5].
Because the stellar density changes with time as the
binary ejects stars, the binary's decay rate can in general
be a complicated function of time.
Two limiting cases are of interest [9].
When the particle number $N$ is small, 
gravitational encounters are able to scatter stars into 
the ``loss cone'' around the binary at a higher rate than
they are ejected.
The density near the binary remains approximately constant
and the decay follows $a^{-1}\sim t$.
This is the ``full loss cone'' regime.
When $N$ is large, 
encounters between stars are weak,
and the binary's loss cone remains nearly empty.
Decay of the binary is limited by the rate at which stars diffuse
into the loss cone; since the diffusion time scales
approximately as $N$, the binary decay follows $a^{-1}\sim t/N$.
This is the ``diffusion'' regime.
Real galaxies are expected to be in the diffusion regime [9].

Figure 1 shows the time evolution of the semi-major axis of the binary
in each of the three integrations.
The binary's energy evolves as an approximately linear function
of the time but with a prefactor that depends  on $N$;
the approximate dependence is
$$
{1\over a} \approx {160 t\over N^{1/3}}.
$$
This is intermediate between the $a^{-1}\propto t$ dependence
of the full loss cone limit, and the $a^{-1}\propto t/N$ dependence
in the diffusion limit.
We conclude that, for the particle numbers considered here,
replenishment of the binary's loss cone is taking place 
but at a lower rate than the rate at which the loss cone is being emptied.
Apparently, particle numbers in excess of $\sim 10^6$ are
required if $N$-body integrations are to be completely in the 
diffusive regime characteristic of real galaxies.
Such large particle numbers can in principle be handled
with direct-summation codes like ours if coupled with
parallel hardware [20].

\figureps[f2.ps,0.7\hsize]{\ 2.}
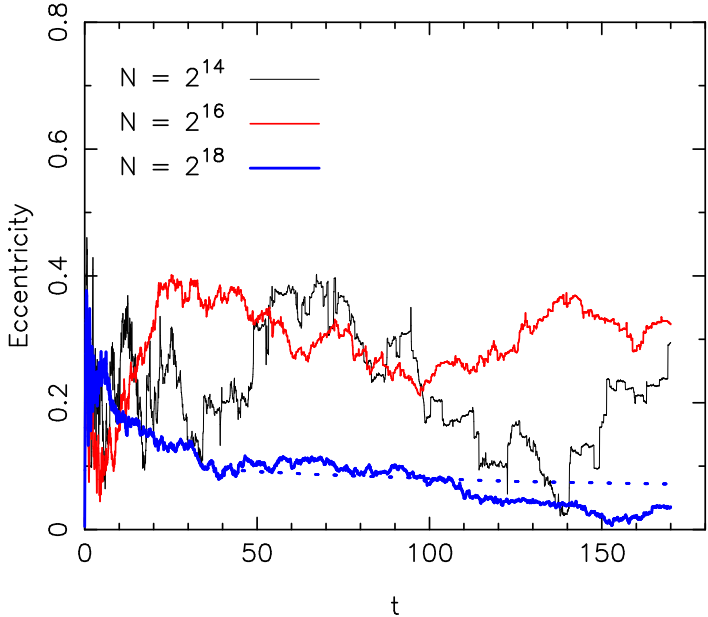
Evolution of the binary orbital eccentricity.

Figure 2 shows the eccentricity evolution of the binary.
The value of $e$ at the time when the hard binary first forms,
$t\approx 30$, is substantially different in the
different integrations due presumably to finite-$N$
effects during the initial inspiral of the two black holes.
Thereafter the eccentricity fluctuates in the case
of the two small-$N$ integrations, but decreases
with time in the integration with the largest $N$.
We compared the eccentricity evolution 
with the predictions of scattering theory.
The rate of change of $e$ is commonly written
$$
{de\over dt} = K {d\over dt} \ln a^{-1}
$$
where $K=K(e,a)$ [4,5].
The functional form of $K(e,a)$ is not well known;
we adopted the expression given in [5] for a hard binary.
We then wrote equation (7) as
$$
e_i = e_{i-1} + K\left(e_{i-1},a_{i-1}\right)\ln\left({a_{i-1}\over a_i}\right)
$$
where the subscript denotes the time step.
Combining equation (8) with the $N$-body
results for $a(t)$, Figure 1, we could then predict
the expected evolution in $e$.
The result for the largest-$N$ integration is shown as 
the dashed line in Figure 2.
There is reasonable agreement, but the eccentricity
evolution even in the largest-$N$ integration still exhibits
substantial fluctuations.
In this regard too, we are not yet in a regime
where the evolution is similar to what would be
expected in real galaxies.

A recent $N$-body study [13] found a much greater degree
of eccentricity evolution, although the initial orbit of the binary
was highly non-circular.

As the binary decays, it ejects stars from the nucleus
and lowers the nuclear density.
Figure 3 shows initial and final density profiles for the
three integrations.
The net effect of a black hole binary on the stellar distribution
is commonly measured in terms of the ``mass deficit,''
defined as the mass in stars that was removed by the binary [2].
We find mass deficits of $(1.94,1.56,1.17)$ in units of the combined
black hole mass in the integrations with 
$N=(0.016,0.065,0.26)\times 10^6$.
These values are of the same order as the mass deficits
inferred in giant elliptical galaxies [2,21,22].

\figureps[f3.ps,0.7\hsize]{\ 3.}
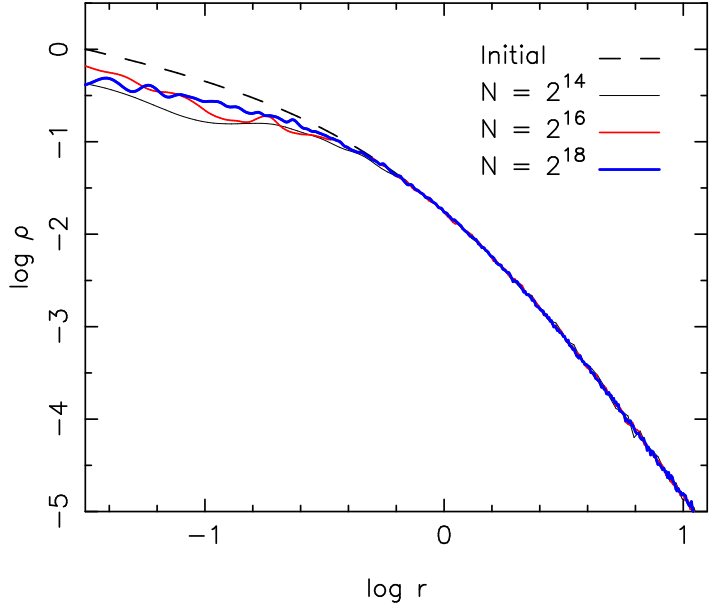
Initial (dashed line) and final density profiles.

\acknowl
This work was supported by grants 
AST-0206031, AST-0420920 and AST-0437519 from the 
NSF, grant 
NNG04GJ48G from NASA,
and grant HST-AR-09519.01-A from
STScI.

\references

\ 1. Thorne, K. S. \& V. B. Braginskii. 1976.
     Gravitational-wave bursts from the nuclei of distant 
     galaxies and quasars - Proposal for detection using 
     Doppler tracking of interplanetary spacecraft.
     Astrophys. J. {\bf 204}: L1-L6.

\ 2. Milosavljevic, M., D. Merritt, A. Rest \& F. van den Bosch. 2002.
     Galaxy cores as relics of black hole mergers.
     Mon. Not. R. Astron. Soc. {\bf 331}: L51-L55.

\ 3. Kandrup, H. E., I. V. Sideris, B. Terzic \& C. L. Bohn. 2003.
     Supermassive black bole binaries as galactic blenders.
     Astrophys. J. {\bf 597}: 111-130.

\ 4. Mikkola, S. \& M. J. Valtonen. 1992.
     Evolution of binaries in the field of light particles and the 
     problem of two black holes.
     Mon. Not. R. Astron. Soc. {\bf 259}: 115-120.

\ 5. Quinlan, G. D. 1996.
     The dynamical evolution of massive black hole binaries
     I. Hardening in a fixed stellar background.
     New Astron. {\bf 1}: 35-56

\ 6. Makino, J. 1997.
     Merging of galaxies with central black holes. II. 
     Evolution of the black hole binary and the structure of the core.
     Astrophys. J. {\bf 478}: 58-65.

\ 7. Milosavljevic, M. \& D. Merritt 2001.
     Formation of galactic nuclei.
     Astrophys. J. {\bf 563}: 34-62.

\ 8. Hemsendorf, M., S. Sigurdsson \& R. Spurzem 2002.
     Collisional dynamics around binary black holes in galactic centers.
     Astrophys. J. {\bf 581}: 1256-1270.

\ 9. Milosavljevic, M. \& D. Merritt. 2003.
     Long-term evolution of massive black hole binaries.
     Astrophys. J. {\bf 596}: 860-878.

\ 10. Makino, J. \& Y. Funato. 2004.
     Evolution of massive black hole binaries.
     Astrophys. J. {\bf 602}: 93-102.

\ 11. Mikkola, S. \& S. J. Aarseth. 1990. 
     A chain regularization method for the few-body problem.
     Celest. Mech. Dyn. Astron. {\bf 47}: 375-390.

\ 12. Mikkola, S. \& S. J.  Aarseth. 1993.
     An implementation of N-body chain regularization.
     Celest. Mech. Dyn. Astron. {\bf 84}: 343-354.

\ 13. Aarseth, S. J. 2003.
      Black hole binary dynamics.
      Astrophys. Sp. Sci. {\bf 285}: 367-372.

\ 14. Mikkola, S. \& S. J.  Aarseth. 2003.
     A time-trasformed leapfrog scheme.
     Celest. Mech. Dyn. Astron. {\bf 57}: 439-459.

\ 15. Preto, M., D. Merritt \& R. Spurzem 2004.
     N-body growth of a Bahcall-Wolf cusp around a black hole.
     Astrophys. J. {\bf 613}: 109-112.

\ 16. Aarseth, S. J. 1999.
      Pub. Astron. Soc. Pac. {\bf 111}: 1333-1346.

\ 17. Aarseth, S. J. 2003.
      Gravitational N-Body Simulations.
      Cambridge University Press. Cambridge.

\ 18. Kustaanheimo, P. \& E. Stiefel. 1965.
      Perturbation theory of Kepler motion based on spinor regularization.
      J. Reine Angew. Math. {\bf 218}: 204-219.

\ 19. W. Dehnen. 1993.
      A family of potential-density pairs for spherical 
      galaxies and bulges.
      Mon. Not. R. Astron. Soc. {\bf 265}: 250-256.

\ 20. Dorband, E. N., M. Hemsendorf \& D. Merritt 2003.
      Systolic and hyper-systolic algorithms for the gravitational
      $N$-body problem, with an application to Brownian motion.
      J. Comp. Phys. {\bf 185}: 484-511.

\ 21. Ravindranath, S., L. C. Ho \& A. V. Filippenko 2002.
      Nuclear cusps and cores in early-type galaxies as relics 
      of binary black hole mergers.
      Astrophys. J. {\bf 566}: 801-808.

\ 22. Graham, A. 2004.
      Core depletion from coalescing supermassive black holes.
      Astrophys. J. {\bf 613}: L33-L36.

\bye